\documentclass[twocolumn,aps,prl,showpacs,floatfix]{revtex4-1}
\usepackage{amsmath}
\usepackage{amsfonts}
\usepackage{amsbsy}
\usepackage{xcolor}
\usepackage{soul}
\usepackage{graphicx}

\newcommand{\ket}[1]{\ensuremath{\left\vert #1 \right\rangle}}

\newcommand{\unitvec}[1]{\hat{\mathbf{{#1}}}}

\newcommand{\levellabel}[1]{\ensuremath{\ket{#1}}}

\newcommand{\bea}{\begin{eqnarray}}
\newcommand{\eea}{\end{eqnarray}}
\newcommand{\beq}{\begin{equation}}
\newcommand{\eeq}{\end{equation}}

\newcommand{\br}{{\bf r}}

\newcommand{\bE}{{\bf E}}

\newcommand{\bd}{{\bf d}}

\newcommand{\pola}{\hat{\mathbf{e}}}
\newcommand{\rv}{{\bf r}}

\newcommand{\Dc}{{\cal D}}

\renewcommand{\>}{\rangle}

\newcommand{\Pc}{\mathcal{P}}

\newcommand{\commentout}[1]{{}}

\newcommand{\cbE}{\boldsymbol{\mathbf{\cal E}}}

\begin{document}
\title{Optical resonance shifts in the fluorescence of thermal and cold atomic gases}
\date{\today}
\author{S. D. Jenkins}
\affiliation{Mathematical Sciences, University of Southampton, Southampton SO17 1BJ, United Kingdom}
\author{J. Ruostekoski}
\affiliation{Mathematical Sciences, University of Southampton, Southampton SO17 1BJ, United Kingdom}
\author{J. Javanainen}
\affiliation{Department of Physics, University of Connecticut, Storrs, Connecticut 06269-3046}
\author{R. Bourgain}
\affiliation{Laboratoire Charles Fabry, Institut d'Optique, CNRS, Univ Paris Sud, 2 Avenue Augustin Fresnel, 91127 Palaiseau cedex, France}
\author{S. Jennewein}
\affiliation{Laboratoire Charles Fabry, Institut d'Optique, CNRS, Univ Paris Sud, 2 Avenue Augustin Fresnel, 91127 Palaiseau cedex, France}
\author{Y. R. P. Sortais}
\affiliation{Laboratoire Charles Fabry, Institut d'Optique, CNRS, Univ Paris Sud, 2 Avenue Augustin Fresnel, 91127 Palaiseau cedex, France}
\author{A. Browaeys}
\affiliation{Laboratoire Charles Fabry, Institut d'Optique, CNRS, Univ Paris Sud, 2 Avenue Augustin Fresnel, 91127 Palaiseau cedex, France}

\begin{abstract}
  We show that the resonance shifts in the fluorescence of a cold gas of rubidium atoms substantially differ
  from those of thermal atomic ensembles that obey the standard continuous medium
  electrodynamics. The analysis is based on large-scale microscopic numerical simulations and experimental measurements of the resonance shifts in a steady-state response in light propagation.
\end{abstract}

\maketitle

An ensemble of resonant emitters can respond strongly to electromagnetic fields. With sufficiently closely-spaced emitters, the radiative response of a single, isolated emitter is no longer a simple guide to the behavior of many. The response of the sample becomes
collective due to strong resonant dipole-dipole (DD) interactions~\cite{Dicke1954,Lehmberg1970a,Lehmberg1970b,Morice1995,
Ruostekoski1997a,Javanainen1999,Berhane2000,Pinheiro2004,Ishimaru1978,vanTiggelen1990,Chomaz2012,
Jenkins2012b,Jenkins2012a,Olmos2013,Castin2013,Javanainen2014,Pellegrino2014,Javanainen2016,Skipetrov2014,Bettles2014}. Owing to improving experimental control, the collective radiative interactions have recently experienced a resurge in interest, both in fundamental studies and in the developments of technological applications. Among the systems investigated are cold atoms~\cite{Balik2013,Bienaime2010,Loew2005,Pellegrino2014,Kemp2014,wilkowski,Ye2016,Guerin_subr16}, thin thermal cells~\cite{Keaveney2012},
photonic crystals~\cite{Segev2013}, metamaterial arrays of nanofabricated resonators~\cite{Lemoult2010,FedotovEtAlPRL2010,Adamo2012}, arrays of ions~\cite{Meir2013}, and nanoemitters~\cite{Brandes2005,Pierrat2010,Diniz2011}. Atoms provide an especially promising system for the studies of collective radiative phenomena, since they make a well-characterized medium with precisely determined radiative resonance frequencies and linewidths, without any true absorption where radiation is lost. Furthermore, cold atomic ensembles form homogeneously broadened systems where the effect of the thermal motion of the atoms on radiative resonance frequencies may be ignored.

Recent numerical simulations~\cite{Javanainen2014} have highlighted how the optical response of cold, dense atomic ensembles can be dramatically different from that of thermal atoms. In cold atomic gases the incident light can induce position-dependent correlations between the atoms due to the light-mediated resonant DD interactions. The thermal motion of hot atoms, in contrast, introduces Doppler shifts in the resonance frequencies of the atoms, which modifies the optical response by suppressing these correlations. With increasing inhomogeneous broadening the atoms are simply farther away from resonance with the light sent by the other atoms, which reduces the light-mediated interactions, as demonstrated in Ref.~\cite{Javanainen2014}.

The standard textbook theory of macroscopic electromagnetism~\cite{Jackson,Born} in a polarizable medium represents an effective-medium mean-field theory (MFT) that assumes each atom interacting with the average behavior of the surrounding atoms. In such models the spatial information about the precise locations of the pointlike atoms -- and the corresponding details of the position-dependent DD interactions -- is washed out, resulting in the absence of the light-induced correlations and in approximations in the calculations of the optical response. In thermal atomic ensembles, at sufficiently high temperatures, the suppression of the DD interactions between the atoms restores the validity of the effective-medium MFT of the standard optics~\cite{Javanainen2014}. In particular, the optical response of thermal atomic gases was found to qualitatively correspond to the low-atom-density limit of the standard optics~\cite{Javanainen2014,Javanainen2016}. Established models of resonance line shifts, the Lorentz-Lorenz (LL) shift and its similarly mean-field theoretical collective (finite-size) counterpart, the
``cooperative Lamb shift''~\cite{Friedberg1973}, have indeed been verified in thin vapor cell experiments on hot atoms~\cite{Keaveney2012}.

Here we compare side by side the resonance shifts measured in cold $^{87}$Rb atomic gases in the low excitation regime with those obtained in large-scale microscopic numerical simulations of cold and hot atomic ensembles. The thermally-induced broadening of hot atoms is generated by stochastically sampling the inhomogeneous broadening of the resonance frequencies of individual atoms. We find that both the experimental observations and the numerical cold-atom calculations of the resonance line shifts substantially deviate from those of thermal atomic ensembles. In particular, in both cases the density-dependent resonance shift is absent. However, introducing inhomogeneous broadening restores the shift.

The numerical simulations incorporate the recurrent scattering processes~\cite{Ishimaru1978} between the atoms where the light is scattered more than once by the same atom. In cold and dense ensembles these lead to strongly sub- and superradiant excitations, and the simulation results demonstrate a range of collective eigenmode decay rates spanning several orders of magnitude. In a hot gas the distribution of the eigenmode decay rates is notably narrower, also indicating the suppression of the DD  interactions as in the MFT effective-medium theories.

As described in Ref.~\cite{Pellegrino2014}, we designed our experimental setup so as to access densities and temperatures at which DD interactions can manifest themselves in the optical response. We laser-cool up to a few hundred $^{87}$Rb atoms in a microscopic dipole trap to obtain an elongated, cigar-shaped cloud at a temperature of $\sim 110\,\mathrm{\mu K}$ \cite{Bourgain2013b}, with root-mean-square sizes of
$\sigma_x\simeq \sigma_y \simeq 0.3 \lambda$ and $\sigma_z \simeq 2.4\lambda$, where $\lambda \simeq 780.2\,\mathrm{nm}$ is the resonant wavelength.
The scattered light intensity is detected in the far field in a direction perpendicular to the propagation of the incident light (fluorescent imaging). The incident light has a low intensity ($I/I_{\rm sat}=0.1$). The control of the atom number $N$ allows the observation of the gradual buildup of the collective radiative response when $N$ is increased~\cite{Pellegrino2014}. The highest atom numbers correspond to peak densities of $\rho \simeq 0.9~k^3$ ($k\equiv2\pi/\lambda$) at the center of the trap, which results in significant DD interactions influencing the optical response of the atoms. At the same time, thermal atomic motion produces only a negligible Doppler broadening of $0.04\gamma$, where
$\gamma = 2\pi \times 3\, \mathrm{MHz}$ [$\gamma = \Dc^2 k^3/ (6\pi\hbar\epsilon_0) $, where $\Dc$ denotes the reduced dipole matrix element] is the half width half maximum linewidth for the studied $|g\>=|5S_{1/2}, F=2 \>\leftrightarrow |e\>=|5P_{3/2},F'=3\>$ transition.

Reference~\cite{Pellegrino2014} reported measurements of light scattering performed by sending a series of light pulses on the atomic cloud, hereafter referred to as the ``burst excitation'' method. Here we report on new experimental protocols based on imaging with a single laser pulse, which rule out potential systematics and check the robustness of the shift measurements of Ref.~\cite{Pellegrino2014}. By a comparison between theory and experiments, we obtain clear evidence of a dramatic difference in the shift of the resonance in light scattering in cold atomic ensembles and the shift predicted for a hot vapor with comparable density.

\begin{figure}
\centering
\includegraphics[width=\columnwidth]{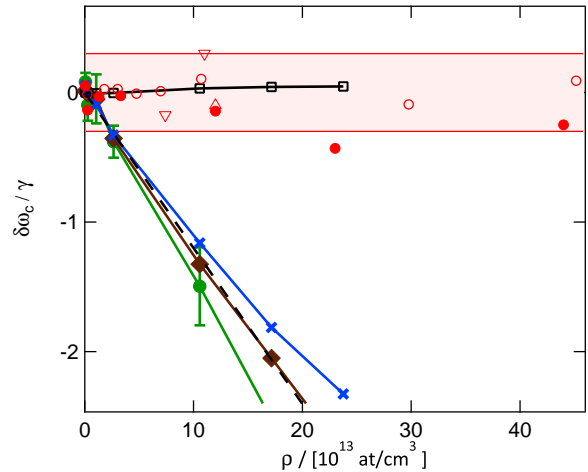}
\vspace{-0.7cm}
\caption{Line shift as a function of the atom density. [Experiments reported in Ref.~\cite{Pellegrino2014}] Filled red circles: excitation with bursts of $200$ light pulses; each point corresponds to a different number of atoms. [Experiments specific to this work] Upper (lower) red triangles: excitation with $75$ ($1$) pulses per burst. Empty red circles: excitation with one pulse after a variable time of flight; the cloud contains $\sim 450$ atoms. The shaded area indicates the laser linewidth of $\pm0.3\gamma$. The densities are known within a factor $2$, due to the cumulated measurement uncertainties on the trap size ($6\%$), atom number ($10\%$) and temperature ($10\%$). Decreasing values of the density correspond to time-of-flights $\Delta t =(0.7, 1.7, 2.7, 3.7, 4.7, 6.7, 8.7, 20.7)\mu$s.
[Simulations] Shift of the line for homogeneously (black empty squares) and inhomogeneously broadened samples with root-mean-square spectral broadening of $10\gamma$ (blue crosses), $20\gamma$ (brown diamonds), and $100\gamma$ (green circles). Error bars: $95\%$ confidence intervals on the shift obtained from the fit of the spectrum to the Voigt profile (see text). Dashed line: estimated Lorentz-Lorenz shift.}
\label{fig:Exp_shift_width}
\end{figure}

In Ref.~\cite{Pellegrino2014} the cloud was excited by a $125$\,ns pulse shortly after the trap was switched off and the atoms were released in free space. The cloud was recaptured in the trap after the excitation was completed, and the same release-excitation-recapture sequence was repeated $200$ times with the same cloud before a new cloud was produced. The free-flight period (after release and before the excitation) was sufficiently short not to affect the atom density. However, the repeated excitation of the same cloud could lead to a possible variation of the effective volume of the atom cloud due to switching on and off the trap and to the small (less than 5\%) parametric heating. The results for the resonance shift are reported in Fig.~\ref{fig:Exp_shift_width}. Each point, for a given atom number, corresponds to an average over typically $1000$ newly-loaded clouds.

To rule out possible systematics due to the repetition of excitation pulses, we performed complementary measurements where we reduced the number of pulses per burst~\footnote{The pulse length is then increased to $700$\,ns to keep the integration time reasonable.}. The results, which we report in Fig.~\ref{fig:Exp_shift_width}, do not indicate any significant change. While this does not entirely exclude the possibility of atom density variation during a single pulse, it does rule out the possible cumulative effect from sending several pulses on the same cloud. Finally, we performed measurements with excitation intensities at even lower levels (down to $I/I_{\rm sat} =0.001$). We still did not see any significant shift in the resonance.

In order to check for the robustness of the absence of the shift in a cold atomic sample, we also implemented a new protocol for the excitation.
Instead of varying the atom number we vary the density of the cloud by changing the geometry of the cloud. After having trapped $\sim 450$ atoms, we switch off the trap and vary the free-flight period $\Delta t$ during which the density of the atoms drops as $N/[(2\pi)^{3/2}\sigma_x \sigma_y \sigma_z]$, with $[\sigma_i(\Delta t)]^2=[\sigma_i(0)]^2+k_B T t^2/m$ ($i=x,y,z$), and the aspect ratio of the cloud evolves from a highly elongated cigar-shaped cloud to a spherical cloud. We then image the atoms with a $2\,\mu$s pulse at a given detuning and repeat the experiment $\sim 1000$ times using a new cloud each time.
The results for the resonance shifts are shown in Fig.~\ref{fig:Exp_shift_width} as a function of the peak density of the cloud at the beginning of the excitation pulse~\footnote{Note that the density drops during the $2\,\mu$s excitation pulse and that this drop is particularly significant for short free-flights.}. The density is deduced from the independent measurements of the trap size, atom number, and temperature of the cloud.
The various experimental protocols were implemented over a period of several months and with numerous adjustments to the experimental apparatus, but the results consistently indicate a very small resonance shift.

In the simulations of the optical response, we consider the weak excitation limit where the saturation of the excited state is ignored.
We include the full internal atomic level structure and the magnetic field level shifts. We stochastically sample the positions of the atoms according to the density distribution as independent identically distributed random variables, so that at each realization we have the $N$ atoms fixed at positions $\br_j$, $j=1,\ldots,N$. Here all the field amplitudes and the atomic polarization correspond to the slowly varying positive frequency components with oscillations at the laser frequency $\omega$.

The atoms initially are in an incoherent mixture of the hyperfine levels with a finite probability $p_m$ of occupying the level $\ket{g,m}$ ($m=-2,\ldots,2$). In the weak excitation limit, this population distribution remains constant during the imaging. For each stochastic realization of fixed atomic positions we similarly sample for each atom $j$ its magnetic Zeeman state, $m_j$ ($j=1,\ldots,N$). The probability of atom $j$ being in state $\ket{g,m}$ is the initial population of that Zeeman state $p_m$ ($0 \le p_m \le 1;\, \sum_m p_m =1$). The optical pumping used in preparation of the ground-state atomic sample skews the initial populations prior to the imaging and we use the experimental estimate of
populations $p_{0}=p_{1}=p_{2}=1/3$ and $p_{-1}=p_{-2}=0$.

Once the atom $j$ is stochastically sampled to be at the position
$\br_j$ and hyperfine state $\nu$, we calculate the dipole moment ${\bf d}_j$ for each atom $j$ when the light is illuminating
the sample. For the multilevel $^{87}$Rb atoms we write $ {\bf d}_j= \Dc
\sum_{\eta,\sigma} \pola_{\sigma} {\cal
  C}_{\nu,\eta}^{(\sigma)}\Pc_{\nu\eta}^{(j)}$.
The summation runs over the unit circular polarization vectors
$\sigma=\pm1,0$ weighted by the Clebsch-Gordan coefficients ${\cal
  C}_{\nu,\eta}^{(\sigma)}$  of the corresponding optical transitions
$|g,\nu\>\rightarrow|e,\nu+\sigma\>=|e,\eta\>$, and $\Pc_{\nu\eta}^{(j)}$ is the atomic excitation amplitude of the transition.
The polarization for the atom in each magnetic sublevel therefore has
three orthogonal vector components.

Each atom $j$ acts as a source of dipole radiation, such that
$\epsilon_0\bE^{(j)}_S(\br)={\sf G}({\bf r}-{\bf r}_j) {\bf d}_j$, where
${\sf G}$ is the dipole radiation kernel and $\bE^{(j)}_S(\br)$
represents the familiar expression of the electric field at $\br$ from
a dipole $\bd_j$  residing at $\br_j$~\cite{Jackson}.
Each of the excitation amplitudes $\Pc_{\nu\eta}^{(j)}$ is then driven
by the sum of the incident field and the fields scattered from all the
other $N-1$ atoms $\bE_{\rm ext}(\br_j) = \cbE_0(\br_j)+\sum_{l\neq j}
\bE^{(l)}_S(\br_j)$. In the steady-state response we have $
\Pc_{\nu\eta}^{(j)} = \alpha_{\nu\eta}\sum_\sigma \pola_{\sigma} {\cal C}_{\nu,\eta}^{(\sigma)} \cdot \epsilon_0 {\bf E}_{\rm ext}(\rv_j)/\Dc
$, where $\alpha_{\nu\eta}=-\Dc^2/[\hbar\epsilon_0 (\Delta_{\nu\eta}+i\gamma)]$
denotes the atomic polarizability.
The detuning from the atomic resonance
$\Delta_{\nu\eta} = \omega-\omega_{\nu\eta}=\omega - \omega_0 + {\mu_BB} (g_e
\eta - g_g\nu)/\hbar$ is given in terms of the Land\'{e} g-factors
$g_g\simeq 0.50$ and
$g_e\simeq 0.67$ for levels $\levellabel{g}$ and $\levellabel{e}$; $\omega_0$ is the resonance frequency of
the $\levellabel{g}\leftrightarrow\levellabel{e}$ transition in the absence of
magnetic field.
Each excitation emits radiation that couples to the excitations of the
other atoms; we obtain a closed set of linear equations that can be
solved to calculate the atomic excitations and then the total electric field $\bE_{\rm}(\br)
= \cbE_0(\br)+\sum_{j} \bE^{(j)}_S(\br)$ everywhere.
We calculate ensemble averages of the scattered light intensity by typically averaging over
several tens or hundreds of thousands of stochastic realizations of atomic
positions and magnetic sublevel configurations. The calculations are done using two sets of independently developed numerical codes.

In order to characterize the differences between the response of cold and thermal atomic ensembles we incorporate the effect of the thermal distribution of the atomic velocities in the simulations. In our simple approach we account for the Maxwell-Boltzmann distribution and the resulting Doppler shifts of the atomic resonances by assigning to each atom a shift of the resonance frequency drawn at random from a Gaussian distribution. In the simulations we consider such inhomogeneous broadenings with root-mean-square thermal widths of 10$\gamma$, 20$\gamma$, and 100$\gamma$. In order to extract the shifts, the calculated spectra are fitted to Voigt profiles that are convolutions of the Lorentzian and Gaussian distributions~\footnote{See Supplemental Material}.

In both experiments and in numerical simulations the optical response is obtained as follows. The atoms respond to an incident field with the polarization $\pola_+=-(\pola_y + i \pola_z) / \sqrt{2}$ propagating antiparallel to the magnetic bias field of $\sim 1 \pola_x$G,  along a tightly-confined radial direction of the trap. The light scattered in the $-z$ direction, along the long axis of the trap, is collected  in the far field by a lens with the numerical aperture $0.5$, and  the signal then passes through a polarizer rotated about $-\unitvec{e}_z$ by $55^\circ$ from the $x$ axis. Finally, the intensity is measured on a CCD camera.

Calculations on few-atom cold ensembles produce the expected Lorentzian line shapes for the spectra of the scattered intensity. As $N$ increases, however, the spectral response begins to deviate from the independent atom scattering, essentially in width and in the amount of scattered light~\cite{Pellegrino2014} but not in the shift of the resonance, which remains small in comparison to the natural linewidth of an individual atom (see Fig.~\ref{fig:Exp_shift_width}). Here, the shift is defined as the difference in the light frequencies that produce the maximum scattered intensity in the given multiatom sample and in a single atom. We find that the calculated shifts are in good agreement with the experimental shifts, which are deduced from Lorentzian fits to the measured spectra, and deviate by less than the linewidth of the laser $\sim \gamma/3$. By contrast, when we introduce the Doppler broadening associated with the thermal Maxwell-Boltzmann distribution of the atomic velocities, we find notably larger shifts. For the Doppler width of $10\gamma$, corresponding to the temperature of $5.5$\,K, the shift is, e.g. at the density $\rho=2.4\times10^{14}$\,cm$^{-3}$, already $50$ times larger than the stationary atom result. Increasing the Doppler broadening further has a weaker effect on the shift. For a hot gas at the same density but with a Doppler width of $100\gamma$, the calculated shift is $-3.3\gamma$.

In continuous effective-medium electrodynamics a natural energy scale for the resonance shifts is the LL shift~\cite{Jackson,Born} $\Delta_{\rm LL} = -2\pi\gamma\rho/k^3$, and at low atom densities $\rho$ one
expects a shift of a resonance $\propto \rho/k^3$ also from dimensional analysis. We may estimate the LL shift by $\rho$ at the center of the trap
(dashed line in Fig.~\ref{fig:Exp_shift_width}). We find that the LL shift is absent both in the experiment and in the electrodynamics simulations of a cold gas. By contrast, introducing inhomogeneous broadening restores a resonance shift that is roughly equal to the LL shift $\Delta_{\rm LL}$, as illustrated in Fig.~\ref{fig:Exp_shift_width}.

\begin{figure}
  \centering
  \includegraphics[width=0.98\columnwidth]{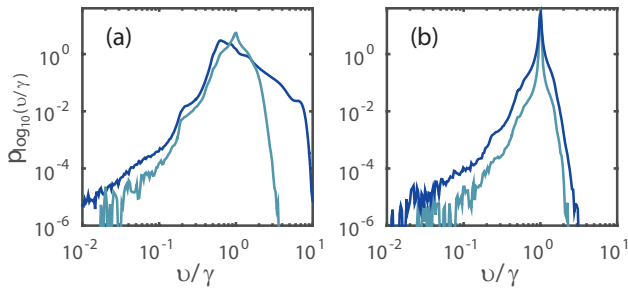}
  \vspace{-0.2cm}
  \caption{Distribution of the logarithm of collective mode decay rates $\upsilon_n$ in a cloud of (a) homogeneously; (b) inhomogeneously; broadened $^{87}$Rb atoms. In (b) the single-atom resonance frequencies have a Gaussian distribution with root-mean-square width $\Delta\omega=100\gamma$. The samples contain $N=50$ (light blue) and $N=450$ (dark blue) atoms, with peak atom densities $\rho=2.6\times10^{13}$\,cm$^{-3}$ and $2.4\times10^{14}$\,cm$^{-3}$, respectively. The initial Zeeman state populations are $p_0=p_1=p_2 = 1/3$ and $p_{-1}=p_{-2}=0$.}
  \label{fig:hist_gamma_hom}
\end{figure}

The effect of strong light-mediated interactions between the atoms can be understood in a collective response of the atomic ensemble where the atoms exhibit collective optical linewidths and line shifts~\cite{Jenkins2012a}. The collective mode characteristics for a particular realization of atomic positions strongly influence the response of the ensemble as a whole. The closer the collective decay rates are to those of a single atom, the better
the scattering dynamics can be approximated by independent atoms, while a broad distribution of decay rates is an indication of strong DD interactions between the atoms.

We calculate the collective eigenmodes for the radiative excitations of the atoms over $51,200$ stochastic realizations of atomic positions. Figure~\ref{fig:hist_gamma_hom} shows the distribution of the decay rates $\upsilon_n$ of the collective modes [from a histogram of the values of $\log_{10}(\upsilon_n)$] for both homogeneously and inhomogenously broadened samples with sub- and superradiant decay rates spanning several orders of magnitude. In the cold samples, the number of atoms in the ensemble strongly influences the breadth of the collective decay rates. For $N=50$ and $N=450$ cold $^{87}$Rb atoms, one percent of collective modes have decay rates of less than $0.45\gamma$ and $0.39\gamma$, respectively, and the median linewidths in these samples are $0.98\gamma$ and $0.79\gamma$, respectively.
The inhibition of the light-mediated interactions in thermal ensembles is also shown in the distribution of the decay rates that is notably narrower. For both $N=50$ and $N=450$ atoms, the median linewidth matches that of a single atom, while one percent of the collective decay rates are below $0.90\gamma$ and $0.62\gamma$, respectively. Overall, we find that increasing the density of cold atoms makes the median value of the linewidth smaller, and generates a long tail of subradiant mode decay rates.

The response of a cold, dense vapor is characterized by the many-atom collective excitation modes. In our case (this generally depends on the geometry of the sample and the excitation protocol~\footnote{The shift can be recovered in the low atom densities or when a specific collective mode that exhibits a shift is driven~\cite{Javanainen2016,Roof16,Araujo16}}) the highly excited modes exhibit resonance frequencies close to the single atom resonance, and the shift in the observed spectrum consequently is small.
In contrast, in thermal ensembles the shift can be described by the standard local-field correction by introducing an exclusion volume~\cite{Jackson,Born} around an independently scattering atom.

In conclusion, we provided side-by-side comparisons of the resonance shifts obtained in fluorescence measurements and in microscopic numerical simulations. We found that the shifts measured in cold atomic gases qualitatively agree with cold-atom simulation results, but substantially differ from those predicted for thermal atomic ensembles. This can be illustrated by progressively increasing the temperature of the atoms in the numerical simulations of our discrete atomic dipole model.

Standard models of macroscopic electromagnetism in a polarizable medium constitute mean-field approximations that ignore the discrete nature of atoms and treat the atomic polarization as a continuous field. Strong DD interactions between closely-spaced atoms can induce correlations between the atoms and large deviations from the MFT models occur surprisingly readily~\cite{Javanainen2016}, even at relatively low densities in optically thin cold samples. This potentially has important consequences on quantum technologies with atom-light interfaces. Here we have illustrated a substantially different behavior of resonance shifts in the fluorescent imaging of trapped, cold Rb atoms from those of thermal atoms.
Parallel to our work, the effects of motional dynamics of atoms were observed in a cold Sr atom vapor by comparing the optical response of narrow and broad linewidth transitions of the atoms~\cite{Ye2016}.

\begin{acknowledgements}
We acknowledge financial support from the Leverhulme Trust, EPSRC, NSF, Grant Nos.\ PHY-0967644 and PHY-1401151, from the E.U.\ through the ERC Starting Grant ARENA
and the HAIRS project, from the Triangle de la Physique (COLISCINA project),
the labex PALM (ECONOMIC project)
and the Region Ile-de-France (LISCOLEM project).
\end{acknowledgements}


\begin{thebibliography}{46}%
\makeatletter
\providecommand \@ifxundefined [1]{%
 \@ifx{#1\undefined}
}%
\providecommand \@ifnum [1]{%
 \ifnum #1\expandafter \@firstoftwo
 \else \expandafter \@secondoftwo
 \fi
}%
\providecommand \@ifx [1]{%
 \ifx #1\expandafter \@firstoftwo
 \else \expandafter \@secondoftwo
 \fi
}%
\providecommand \natexlab [1]{#1}%
\providecommand \enquote  [1]{``#1''}%
\providecommand \bibnamefont  [1]{#1}%
\providecommand \bibfnamefont [1]{#1}%
\providecommand \citenamefont [1]{#1}%
\providecommand \href@noop [0]{\@secondoftwo}%
\providecommand \href [0]{\begingroup \@sanitize@url \@href}%
\providecommand \@href[1]{\@@startlink{#1}\@@href}%
\providecommand \@@href[1]{\endgroup#1\@@endlink}%
\providecommand \@sanitize@url [0]{\catcode `\\12\catcode `\$12\catcode
  `\&12\catcode `\#12\catcode `\^12\catcode `\_12\catcode `\%12\relax}%
\providecommand \@@startlink[1]{}%
\providecommand \@@endlink[0]{}%
\providecommand \url  [0]{\begingroup\@sanitize@url \@url }%
\providecommand \@url [1]{\endgroup\@href {#1}{\urlprefix }}%
\providecommand \urlprefix  [0]{URL }%
\providecommand \Eprint [0]{\href }%
\providecommand \doibase [0]{http://dx.doi.org/}%
\providecommand \selectlanguage [0]{\@gobble}%
\providecommand \bibinfo  [0]{\@secondoftwo}%
\providecommand \bibfield  [0]{\@secondoftwo}%
\providecommand \translation [1]{[#1]}%
\providecommand \BibitemOpen [0]{}%
\providecommand \bibitemStop [0]{}%
\providecommand \bibitemNoStop [0]{.\EOS\space}%
\providecommand \EOS [0]{\spacefactor3000\relax}%
\providecommand \BibitemShut  [1]{\csname bibitem#1\endcsname}%
\let\auto@bib@innerbib\@empty
\bibitem [{\citenamefont {Dicke}(1954)}]{Dicke1954}%
  \BibitemOpen
  \bibfield  {author} {\bibinfo {author} {\bibfnamefont {R.~H.}\ \bibnamefont
  {Dicke}},\ }\href {\doibase 10.1103/PhysRev.93.99} {\bibfield  {journal}
  {\bibinfo  {journal} {Phys. Rev.}\ }\textbf {\bibinfo {volume} {93}},\
  \bibinfo {pages} {99} (\bibinfo {year} {1954})}\BibitemShut {NoStop}%
\bibitem [{\citenamefont {Lehmberg}(1970{\natexlab{a}})}]{Lehmberg1970a}%
  \BibitemOpen
  \bibfield  {author} {\bibinfo {author} {\bibfnamefont {R.~H.}\ \bibnamefont
  {Lehmberg}},\ }\href {\doibase 10.1103/PhysRevA.2.883} {\bibfield  {journal}
  {\bibinfo  {journal} {Phys. Rev. A}\ }\textbf {\bibinfo {volume} {2}},\
  \bibinfo {pages} {883} (\bibinfo {year} {1970}{\natexlab{a}})}\BibitemShut
  {NoStop}%
\bibitem [{\citenamefont {Lehmberg}(1970{\natexlab{b}})}]{Lehmberg1970b}%
  \BibitemOpen
  \bibfield  {author} {\bibinfo {author} {\bibfnamefont {R.~H.}\ \bibnamefont
  {Lehmberg}},\ }\href {\doibase 10.1103/PhysRevA.2.889} {\bibfield  {journal}
  {\bibinfo  {journal} {Phys. Rev. A}\ }\textbf {\bibinfo {volume} {2}},\
  \bibinfo {pages} {889} (\bibinfo {year} {1970}{\natexlab{b}})}\BibitemShut
  {NoStop}%
\bibitem [{\citenamefont {Morice}\ \emph {et~al.}(1995)\citenamefont {Morice},
  \citenamefont {Castin},\ and\ \citenamefont {Dalibard}}]{Morice1995}%
  \BibitemOpen
  \bibfield  {author} {\bibinfo {author} {\bibfnamefont {O.}~\bibnamefont
  {Morice}}, \bibinfo {author} {\bibfnamefont {Y.}~\bibnamefont {Castin}}, \
  and\ \bibinfo {author} {\bibfnamefont {J.}~\bibnamefont {Dalibard}},\
  }\href@noop {} {\bibfield  {journal} {\bibinfo  {journal} {Phys. Rev. A}\
  }\textbf {\bibinfo {volume} {51}},\ \bibinfo {pages} {3896} (\bibinfo {year}
  {1995})}\BibitemShut {NoStop}%
\bibitem [{\citenamefont {Ruostekoski}\ and\ \citenamefont
  {Javanainen}(1997)}]{Ruostekoski1997a}%
  \BibitemOpen
  \bibfield  {author} {\bibinfo {author} {\bibfnamefont {J.}~\bibnamefont
  {Ruostekoski}}\ and\ \bibinfo {author} {\bibfnamefont {J.}~\bibnamefont
  {Javanainen}},\ }\href@noop {} {\bibfield  {journal} {\bibinfo  {journal}
  {Phys. Rev. A}\ }\textbf {\bibinfo {volume} {55}},\ \bibinfo {pages} {513}
  (\bibinfo {year} {1997})}\BibitemShut {NoStop}%
\bibitem [{\citenamefont {Javanainen}\ \emph {et~al.}(1999)\citenamefont
  {Javanainen}, \citenamefont {Ruostekoski}, \citenamefont {Vestergaard},\ and\
  \citenamefont {Francis}}]{Javanainen1999}%
  \BibitemOpen
  \bibfield  {author} {\bibinfo {author} {\bibfnamefont {J.}~\bibnamefont
  {Javanainen}}, \bibinfo {author} {\bibfnamefont {J.}~\bibnamefont
  {Ruostekoski}}, \bibinfo {author} {\bibfnamefont {B.}~\bibnamefont
  {Vestergaard}}, \ and\ \bibinfo {author} {\bibfnamefont {M.~R.}\ \bibnamefont
  {Francis}},\ }\href@noop {} {\bibfield  {journal} {\bibinfo  {journal} {Phys.
  Rev. A}\ }\textbf {\bibinfo {volume} {59}},\ \bibinfo {pages} {649} (\bibinfo
  {year} {1999})}\BibitemShut {NoStop}%
\bibitem [{\citenamefont {Berhane}\ and\ \citenamefont
  {Kennedy}(2000)}]{Berhane2000}%
  \BibitemOpen
  \bibfield  {author} {\bibinfo {author} {\bibfnamefont {B.}~\bibnamefont
  {Berhane}}\ and\ \bibinfo {author} {\bibfnamefont {T.~A.~B.}\ \bibnamefont
  {Kennedy}},\ }\href {\doibase 10.1103/PhysRevA.62.033611} {\bibfield
  {journal} {\bibinfo  {journal} {Phys. Rev. A}\ }\textbf {\bibinfo {volume}
  {62}},\ \bibinfo {pages} {033611} (\bibinfo {year} {2000})}\BibitemShut
  {NoStop}%
\bibitem [{\citenamefont {Pinheiro}\ \emph {et~al.}(2004)\citenamefont
  {Pinheiro}, \citenamefont {Rusek}, \citenamefont {Orlowski},\ and\
  \citenamefont {van Tiggelen}}]{Pinheiro2004}%
  \BibitemOpen
  \bibfield  {author} {\bibinfo {author} {\bibfnamefont {F.~A.}\ \bibnamefont
  {Pinheiro}}, \bibinfo {author} {\bibfnamefont {M.}~\bibnamefont {Rusek}},
  \bibinfo {author} {\bibfnamefont {A.}~\bibnamefont {Orlowski}}, \ and\
  \bibinfo {author} {\bibfnamefont {B.~A.}\ \bibnamefont {van Tiggelen}},\
  }\href {\doibase 10.1103/PhysRevE.69.026605} {\bibfield  {journal} {\bibinfo
  {journal} {Phys. Rev. E}\ }\textbf {\bibinfo {volume} {69}},\ \bibinfo
  {pages} {026605} (\bibinfo {year} {2004})}\BibitemShut {NoStop}%
\bibitem [{\citenamefont {Ishimaru}(1978)}]{Ishimaru1978}%
  \BibitemOpen
  \bibfield  {author} {\bibinfo {author} {\bibfnamefont {A.}~\bibnamefont
  {Ishimaru}},\ }\href@noop {} {\emph {\bibinfo {title} {Wave Propagation and
  Scattering in Random Media: Multiple Scattering, Turbulence, Rough Surfaces,
  and Remote-Sensing}}},\ Vol.~\bibinfo {volume} {2}\ (\bibinfo  {publisher}
  {Academic Press},\ \bibinfo {address} {St. Louis, Missouri},\ \bibinfo {year}
  {1978})\BibitemShut {NoStop}%
\bibitem [{\citenamefont {van Tiggelen}\ \emph {et~al.}(1990)\citenamefont {van
  Tiggelen}, \citenamefont {Lagendijk},\ and\ \citenamefont
  {Tip}}]{vanTiggelen1990}%
  \BibitemOpen
  \bibfield  {author} {\bibinfo {author} {\bibfnamefont {B.}~\bibnamefont {van
  Tiggelen}}, \bibinfo {author} {\bibfnamefont {A.}~\bibnamefont {Lagendijk}},
  \ and\ \bibinfo {author} {\bibfnamefont {A.}~\bibnamefont {Tip}},\
  }\href@noop {} {\bibfield  {journal} {\bibinfo  {journal} {J. Phys. Cond.
  Mat.}\ }\textbf {\bibinfo {volume} {2}},\ \bibinfo {pages} {7653} (\bibinfo
  {year} {1990})}\BibitemShut {NoStop}%
\bibitem [{\citenamefont {Chomaz}\ \emph {et~al.}(2012)\citenamefont {Chomaz},
  \citenamefont {Corman}, \citenamefont {Yefsah}, \citenamefont {Desbuquois},\
  and\ \citenamefont {Dalibard}}]{Chomaz2012}%
  \BibitemOpen
  \bibfield  {author} {\bibinfo {author} {\bibfnamefont {L.}~\bibnamefont
  {Chomaz}}, \bibinfo {author} {\bibfnamefont {L.}~\bibnamefont {Corman}},
  \bibinfo {author} {\bibfnamefont {T.}~\bibnamefont {Yefsah}}, \bibinfo
  {author} {\bibfnamefont {R.}~\bibnamefont {Desbuquois}}, \ and\ \bibinfo
  {author} {\bibfnamefont {J.}~\bibnamefont {Dalibard}},\ }\href@noop {}
  {\bibfield  {journal} {\bibinfo  {journal} {New Journal of Physics}\ }\textbf
  {\bibinfo {volume} {14}},\ \bibinfo {pages} {055001} (\bibinfo {year}
  {2012})}\BibitemShut {NoStop}%
\bibitem [{\citenamefont {Jenkins}\ and\ \citenamefont
  {Ruostekoski}(2012{\natexlab{a}})}]{Jenkins2012b}%
  \BibitemOpen
  \bibfield  {author} {\bibinfo {author} {\bibfnamefont {S.~D.}\ \bibnamefont
  {Jenkins}}\ and\ \bibinfo {author} {\bibfnamefont {J.}~\bibnamefont
  {Ruostekoski}},\ }\href@noop {} {\bibfield  {journal} {\bibinfo  {journal}
  {Phys. Rev. B}\ }\textbf {\bibinfo {volume} {86}},\ \bibinfo {pages} {085116}
  (\bibinfo {year} {2012}{\natexlab{a}})}\BibitemShut {NoStop}%
\bibitem [{\citenamefont {Jenkins}\ and\ \citenamefont
  {Ruostekoski}(2012{\natexlab{b}})}]{Jenkins2012a}%
  \BibitemOpen
  \bibfield  {author} {\bibinfo {author} {\bibfnamefont {S.~D.}\ \bibnamefont
  {Jenkins}}\ and\ \bibinfo {author} {\bibfnamefont {J.}~\bibnamefont
  {Ruostekoski}},\ }\href@noop {} {\bibfield  {journal} {\bibinfo  {journal}
  {Phys. Rev. A}\ }\textbf {\bibinfo {volume} {86}},\ \bibinfo {pages}
  {031602(R)} (\bibinfo {year} {2012}{\natexlab{b}})}\BibitemShut {NoStop}%
\bibitem [{\citenamefont {Olmos}\ \emph {et~al.}(2013)\citenamefont {Olmos},
  \citenamefont {Yu}, \citenamefont {Singh}, \citenamefont {Schreck},
  \citenamefont {Bongs},\ and\ \citenamefont {Lesanovsky}}]{Olmos2013}%
  \BibitemOpen
  \bibfield  {author} {\bibinfo {author} {\bibfnamefont {B.}~\bibnamefont
  {Olmos}}, \bibinfo {author} {\bibfnamefont {D.}~\bibnamefont {Yu}}, \bibinfo
  {author} {\bibfnamefont {Y.}~\bibnamefont {Singh}}, \bibinfo {author}
  {\bibfnamefont {F.}~\bibnamefont {Schreck}}, \bibinfo {author} {\bibfnamefont
  {K.}~\bibnamefont {Bongs}}, \ and\ \bibinfo {author} {\bibfnamefont
  {I.}~\bibnamefont {Lesanovsky}},\ }\href {\doibase
  10.1103/PhysRevLett.110.143602} {\bibfield  {journal} {\bibinfo  {journal}
  {Phys. Rev. Lett.}\ }\textbf {\bibinfo {volume} {110}},\ \bibinfo {pages}
  {143602} (\bibinfo {year} {2013})}\BibitemShut {NoStop}%
\bibitem [{\citenamefont {Antezza}\ and\ \citenamefont
  {Castin}(2013)}]{Castin2013}%
  \BibitemOpen
  \bibfield  {author} {\bibinfo {author} {\bibfnamefont {M.}~\bibnamefont
  {Antezza}}\ and\ \bibinfo {author} {\bibfnamefont {Y.}~\bibnamefont
  {Castin}},\ }\href {\doibase 10.1103/PhysRevA.88.033844} {\bibfield
  {journal} {\bibinfo  {journal} {Phys. Rev. A}\ }\textbf {\bibinfo {volume}
  {88}},\ \bibinfo {pages} {033844} (\bibinfo {year} {2013})}\BibitemShut
  {NoStop}%
\bibitem [{\citenamefont {Javanainen}\ \emph {et~al.}(2014)\citenamefont
  {Javanainen}, \citenamefont {Ruostekoski}, \citenamefont {Li},\ and\
  \citenamefont {Yoo}}]{Javanainen2014}%
  \BibitemOpen
  \bibfield  {author} {\bibinfo {author} {\bibfnamefont {J.}~\bibnamefont
  {Javanainen}}, \bibinfo {author} {\bibfnamefont {J.}~\bibnamefont
  {Ruostekoski}}, \bibinfo {author} {\bibfnamefont {Y.}~\bibnamefont {Li}}, \
  and\ \bibinfo {author} {\bibfnamefont {S.-M.}\ \bibnamefont {Yoo}},\ }\href
  {\doibase 10.1103/PhysRevLett.112.113603} {\bibfield  {journal} {\bibinfo
  {journal} {Phys. Rev. Lett.}\ }\textbf {\bibinfo {volume} {112}},\ \bibinfo
  {pages} {113603} (\bibinfo {year} {2014})}\BibitemShut {NoStop}%
\bibitem [{\citenamefont {Pellegrino}\ \emph {et~al.}(2014)\citenamefont
  {Pellegrino}, \citenamefont {Bourgain}, \citenamefont {Jennewein},
  \citenamefont {Sortais}, \citenamefont {Browaeys}, \citenamefont {Jenkins},\
  and\ \citenamefont {Ruostekoski}}]{Pellegrino2014}%
  \BibitemOpen
  \bibfield  {author} {\bibinfo {author} {\bibfnamefont {J.}~\bibnamefont
  {Pellegrino}}, \bibinfo {author} {\bibfnamefont {R.}~\bibnamefont
  {Bourgain}}, \bibinfo {author} {\bibfnamefont {S.}~\bibnamefont {Jennewein}},
  \bibinfo {author} {\bibfnamefont {Y.~R.~P.}\ \bibnamefont {Sortais}},
  \bibinfo {author} {\bibfnamefont {A.}~\bibnamefont {Browaeys}}, \bibinfo
  {author} {\bibfnamefont {S.~D.}\ \bibnamefont {Jenkins}}, \ and\ \bibinfo
  {author} {\bibfnamefont {J.}~\bibnamefont {Ruostekoski}},\ }\href@noop {}
  {\bibfield  {journal} {\bibinfo  {journal} {Phys. Rev. Lett.}\ }\textbf
  {\bibinfo {volume} {113}},\ \bibinfo {pages} {133602} (\bibinfo {year}
  {2014})}\BibitemShut {NoStop}%
\bibitem [{\citenamefont {Javanainen}\ and\ \citenamefont
  {Ruostekoski}(2016)}]{Javanainen2016}%
  \BibitemOpen
  \bibfield  {author} {\bibinfo {author} {\bibfnamefont {J.}~\bibnamefont
  {Javanainen}}\ and\ \bibinfo {author} {\bibfnamefont {J.}~\bibnamefont
  {Ruostekoski}},\ }\href {\doibase 10.1364/OE.24.000993} {\bibfield  {journal}
  {\bibinfo  {journal} {Opt. Express}\ }\textbf {\bibinfo {volume} {24}},\
  \bibinfo {pages} {993} (\bibinfo {year} {2016})}\BibitemShut {NoStop}%
\bibitem [{\citenamefont {Skipetrov}\ and\ \citenamefont
  {Sokolov}(2014)}]{Skipetrov2014}%
  \BibitemOpen
  \bibfield  {author} {\bibinfo {author} {\bibfnamefont {S.~E.}\ \bibnamefont
  {Skipetrov}}\ and\ \bibinfo {author} {\bibfnamefont {I.~M.}\ \bibnamefont
  {Sokolov}},\ }\href {\doibase 10.1103/PhysRevLett.112.023905} {\bibfield
  {journal} {\bibinfo  {journal} {Phys. Rev. Lett.}\ }\textbf {\bibinfo
  {volume} {112}},\ \bibinfo {pages} {023905} (\bibinfo {year}
  {2014})}\BibitemShut {NoStop}%
\bibitem [{\citenamefont {Bettles}\ \emph {et~al.}(2015)\citenamefont
  {Bettles}, \citenamefont {Gardiner},\ and\ \citenamefont
  {Adams}}]{Bettles2014}%
  \BibitemOpen
  \bibfield  {author} {\bibinfo {author} {\bibfnamefont {R.~J.}\ \bibnamefont
  {Bettles}}, \bibinfo {author} {\bibfnamefont {S.~A.}\ \bibnamefont
  {Gardiner}}, \ and\ \bibinfo {author} {\bibfnamefont {C.~S.}\ \bibnamefont
  {Adams}},\ }\href {\doibase 10.1103/PhysRevA.92.063822} {\bibfield  {journal}
  {\bibinfo  {journal} {Phys. Rev. A}\ }\textbf {\bibinfo {volume} {92}},\
  \bibinfo {pages} {063822} (\bibinfo {year} {2015})}\BibitemShut {NoStop}%
\bibitem [{\citenamefont {Balik}\ \emph {et~al.}(2013)\citenamefont {Balik},
  \citenamefont {Win}, \citenamefont {Havey}, \citenamefont {Sokolov},\ and\
  \citenamefont {Kupriyanov}}]{Balik2013}%
  \BibitemOpen
  \bibfield  {author} {\bibinfo {author} {\bibfnamefont {S.}~\bibnamefont
  {Balik}}, \bibinfo {author} {\bibfnamefont {A.~L.}\ \bibnamefont {Win}},
  \bibinfo {author} {\bibfnamefont {M.~D.}\ \bibnamefont {Havey}}, \bibinfo
  {author} {\bibfnamefont {I.~M.}\ \bibnamefont {Sokolov}}, \ and\ \bibinfo
  {author} {\bibfnamefont {D.~V.}\ \bibnamefont {Kupriyanov}},\ }\href@noop {}
  {\bibfield  {journal} {\bibinfo  {journal} {Phys. Rev. A}\ }\textbf {\bibinfo
  {volume} {87}},\ \bibinfo {pages} {053817} (\bibinfo {year}
  {2013})}\BibitemShut {NoStop}%
\bibitem [{\citenamefont {Bienaim{\'e}}\ \emph {et~al.}(2010)\citenamefont
  {Bienaim{\'e}}, \citenamefont {Bux}, \citenamefont {Lucioni}, \citenamefont
  {Courteille}, \citenamefont {Piovella},\ and\ \citenamefont
  {Kaiser}}]{Bienaime2010}%
  \BibitemOpen
  \bibfield  {author} {\bibinfo {author} {\bibfnamefont {T.}~\bibnamefont
  {Bienaim{\'e}}}, \bibinfo {author} {\bibfnamefont {S.}~\bibnamefont {Bux}},
  \bibinfo {author} {\bibfnamefont {E.}~\bibnamefont {Lucioni}}, \bibinfo
  {author} {\bibfnamefont {P.~W.}\ \bibnamefont {Courteille}}, \bibinfo
  {author} {\bibfnamefont {N.}~\bibnamefont {Piovella}}, \ and\ \bibinfo
  {author} {\bibfnamefont {R.}~\bibnamefont {Kaiser}},\ }\href@noop {}
  {\bibfield  {journal} {\bibinfo  {journal} {Phys. Rev. Lett.}\ }\textbf
  {\bibinfo {volume} {104}},\ \bibinfo {pages} {183602} (\bibinfo {year}
  {2010})}\BibitemShut {NoStop}%
\bibitem [{\citenamefont {L\"o{}w}\ \emph {et~al.}(2005)\citenamefont
  {L\"o{}w}, \citenamefont {Gati}, \citenamefont {Stuhler},\ and\ \citenamefont
  {Pfau}}]{Loew2005}%
  \BibitemOpen
  \bibfield  {author} {\bibinfo {author} {\bibfnamefont {R.}~\bibnamefont
  {L\"o{}w}}, \bibinfo {author} {\bibfnamefont {R.}~\bibnamefont {Gati}},
  \bibinfo {author} {\bibfnamefont {J.}~\bibnamefont {Stuhler}}, \ and\
  \bibinfo {author} {\bibfnamefont {T.}~\bibnamefont {Pfau}},\ }\href@noop {}
  {\bibfield  {journal} {\bibinfo  {journal} {Europhys. Lett.}\ }\textbf
  {\bibinfo {volume} {71}},\ \bibinfo {pages} {214} (\bibinfo {year}
  {2005})}\BibitemShut {NoStop}%
\bibitem [{\citenamefont {Kemp}\ \emph {et~al.}(2014)\citenamefont {Kemp},
  \citenamefont {Roof}, \citenamefont {Havey}, \citenamefont {Sokolov},\ and\
  \citenamefont {Kupriyanov}}]{Kemp2014}%
  \BibitemOpen
  \bibfield  {author} {\bibinfo {author} {\bibfnamefont {K.}~\bibnamefont
  {Kemp}}, \bibinfo {author} {\bibfnamefont {S.~J.}\ \bibnamefont {Roof}},
  \bibinfo {author} {\bibfnamefont {M.~D.}\ \bibnamefont {Havey}}, \bibinfo
  {author} {\bibfnamefont {I.~M.}\ \bibnamefont {Sokolov}}, \ and\ \bibinfo
  {author} {\bibfnamefont {D.~V.}\ \bibnamefont {Kupriyanov}},\ }\href@noop {}
  {\bibfield  {journal} {\bibinfo  {journal} {http://arxiv.org/abs/1410.2497}\
  } (\bibinfo {year} {2014})}\BibitemShut {NoStop}%
\bibitem [{\citenamefont {Kwong}\ \emph {et~al.}(2014)\citenamefont {Kwong},
  \citenamefont {Yang}, \citenamefont {Pramod}, \citenamefont {Pandey},
  \citenamefont {Delande}, \citenamefont {Pierrat},\ and\ \citenamefont
  {Wilkowski}}]{wilkowski}%
  \BibitemOpen
  \bibfield  {author} {\bibinfo {author} {\bibfnamefont {C.~C.}\ \bibnamefont
  {Kwong}}, \bibinfo {author} {\bibfnamefont {T.}~\bibnamefont {Yang}},
  \bibinfo {author} {\bibfnamefont {M.~S.}\ \bibnamefont {Pramod}}, \bibinfo
  {author} {\bibfnamefont {K.}~\bibnamefont {Pandey}}, \bibinfo {author}
  {\bibfnamefont {D.}~\bibnamefont {Delande}}, \bibinfo {author} {\bibfnamefont
  {R.}~\bibnamefont {Pierrat}}, \ and\ \bibinfo {author} {\bibfnamefont
  {D.}~\bibnamefont {Wilkowski}},\ }\href {\doibase
  10.1103/PhysRevLett.113.223601} {\bibfield  {journal} {\bibinfo  {journal}
  {Phys. Rev. Lett.}\ }\textbf {\bibinfo {volume} {113}},\ \bibinfo {pages}
  {223601} (\bibinfo {year} {2014})}\BibitemShut {NoStop}%
\bibitem [{\citenamefont {Bromley}\ \emph {et~al.}(2016)\citenamefont
  {Bromley}, \citenamefont {Zhu}, \citenamefont {Bishof}, \citenamefont
  {Zhang}, \citenamefont {Bothwell}, \citenamefont {Schachenmayer},
  \citenamefont {Nicholson}, \citenamefont {Kaiser}, \citenamefont {Yelin},
  \citenamefont {Lukin}, \citenamefont {Rey},\ and\ \citenamefont
  {Ye}}]{Ye2016}%
  \BibitemOpen
  \bibfield  {author} {\bibinfo {author} {\bibfnamefont {S.~L.}\ \bibnamefont
  {Bromley}}, \bibinfo {author} {\bibfnamefont {B.}~\bibnamefont {Zhu}},
  \bibinfo {author} {\bibfnamefont {M.}~\bibnamefont {Bishof}}, \bibinfo
  {author} {\bibfnamefont {X.}~\bibnamefont {Zhang}}, \bibinfo {author}
  {\bibfnamefont {T.}~\bibnamefont {Bothwell}}, \bibinfo {author}
  {\bibfnamefont {J.}~\bibnamefont {Schachenmayer}}, \bibinfo {author}
  {\bibfnamefont {T.~L.}\ \bibnamefont {Nicholson}}, \bibinfo {author}
  {\bibfnamefont {R.}~\bibnamefont {Kaiser}}, \bibinfo {author} {\bibfnamefont
  {S.~F.}\ \bibnamefont {Yelin}}, \bibinfo {author} {\bibfnamefont {M.~D.}\
  \bibnamefont {Lukin}}, \bibinfo {author} {\bibfnamefont {A.~M.}\ \bibnamefont
  {Rey}}, \ and\ \bibinfo {author} {\bibfnamefont {J.}~\bibnamefont {Ye}},\
  }\href {http://dx.doi.org/10.1038/ncomms11039} {\bibfield  {journal}
  {\bibinfo  {journal} {Nat Commun}\ }\textbf {\bibinfo {volume} {7}},\
  \bibinfo {pages} {11039} (\bibinfo {year} {2016})}\BibitemShut {NoStop}%
\bibitem [{\citenamefont {Guerin}\ \emph {et~al.}(2016)\citenamefont {Guerin},
  \citenamefont {Ara\'ujo},\ and\ \citenamefont {Kaiser}}]{Guerin_subr16}%
  \BibitemOpen
  \bibfield  {author} {\bibinfo {author} {\bibfnamefont {W.}~\bibnamefont
  {Guerin}}, \bibinfo {author} {\bibfnamefont {M.~O.}\ \bibnamefont
  {Ara\'ujo}}, \ and\ \bibinfo {author} {\bibfnamefont {R.}~\bibnamefont
  {Kaiser}},\ }\href {\doibase 10.1103/PhysRevLett.116.083601} {\bibfield
  {journal} {\bibinfo  {journal} {Phys. Rev. Lett.}\ }\textbf {\bibinfo
  {volume} {116}},\ \bibinfo {pages} {083601} (\bibinfo {year}
  {2016})}\BibitemShut {NoStop}%
\bibitem [{\citenamefont {Keaveney}\ \emph {et~al.}(2012)\citenamefont
  {Keaveney}, \citenamefont {Sargsyan}, \citenamefont {Krohn}, \citenamefont
  {Hughes}, \citenamefont {Sarkisyan},\ and\ \citenamefont
  {Adams}}]{Keaveney2012}%
  \BibitemOpen
  \bibfield  {author} {\bibinfo {author} {\bibfnamefont {J.}~\bibnamefont
  {Keaveney}}, \bibinfo {author} {\bibfnamefont {A.}~\bibnamefont {Sargsyan}},
  \bibinfo {author} {\bibfnamefont {U.}~\bibnamefont {Krohn}}, \bibinfo
  {author} {\bibfnamefont {I.~G.}\ \bibnamefont {Hughes}}, \bibinfo {author}
  {\bibfnamefont {D.}~\bibnamefont {Sarkisyan}}, \ and\ \bibinfo {author}
  {\bibfnamefont {C.~S.}\ \bibnamefont {Adams}},\ }\href@noop {} {\bibfield
  {journal} {\bibinfo  {journal} {Phys. Rev. Lett.}\ }\textbf {\bibinfo
  {volume} {108}},\ \bibinfo {pages} {173601} (\bibinfo {year}
  {2012})}\BibitemShut {NoStop}%
\bibitem [{\citenamefont {Segev}\ \emph {et~al.}(2013)\citenamefont {Segev},
  \citenamefont {Silberberg},\ and\ \citenamefont
  {Christodoulides}}]{Segev2013}%
  \BibitemOpen
  \bibfield  {author} {\bibinfo {author} {\bibfnamefont {M.}~\bibnamefont
  {Segev}}, \bibinfo {author} {\bibfnamefont {Y.}~\bibnamefont {Silberberg}}, \
  and\ \bibinfo {author} {\bibfnamefont {D.~N.}\ \bibnamefont
  {Christodoulides}},\ }\href@noop {} {\bibfield  {journal} {\bibinfo
  {journal} {Nat. Phot.}\ }\textbf {\bibinfo {volume} {7}},\ \bibinfo {pages}
  {197} (\bibinfo {year} {2013})}\BibitemShut {NoStop}%
\bibitem [{\citenamefont {Lemoult}\ \emph {et~al.}(2010)\citenamefont
  {Lemoult}, \citenamefont {Lerosey}, \citenamefont {{de Rosny}},\ and\
  \citenamefont {Fink}}]{Lemoult2010}%
  \BibitemOpen
  \bibfield  {author} {\bibinfo {author} {\bibfnamefont {F.}~\bibnamefont
  {Lemoult}}, \bibinfo {author} {\bibfnamefont {G.}~\bibnamefont {Lerosey}},
  \bibinfo {author} {\bibfnamefont {J.}~\bibnamefont {{de Rosny}}}, \ and\
  \bibinfo {author} {\bibfnamefont {M.}~\bibnamefont {Fink}},\ }\href {\doibase
  10.1103/PhysRevLett.104.203901} {\bibfield  {journal} {\bibinfo  {journal}
  {Phys. Rev. Lett.}\ }\textbf {\bibinfo {volume} {104}},\ \bibinfo {pages}
  {203901} (\bibinfo {year} {2010})}\BibitemShut {NoStop}%
\bibitem [{\citenamefont {Fedotov}\ \emph {et~al.}(2010)\citenamefont
  {Fedotov}, \citenamefont {Papasimakis}, \citenamefont {Plum}, \citenamefont
  {Bitzer}, \citenamefont {Walther}, \citenamefont {Kuo}, \citenamefont
  {Tsai},\ and\ \citenamefont {Zheludev}}]{FedotovEtAlPRL2010}%
  \BibitemOpen
  \bibfield  {author} {\bibinfo {author} {\bibfnamefont {V.~A.}\ \bibnamefont
  {Fedotov}}, \bibinfo {author} {\bibfnamefont {N.}~\bibnamefont
  {Papasimakis}}, \bibinfo {author} {\bibfnamefont {E.}~\bibnamefont {Plum}},
  \bibinfo {author} {\bibfnamefont {A.}~\bibnamefont {Bitzer}}, \bibinfo
  {author} {\bibfnamefont {M.}~\bibnamefont {Walther}}, \bibinfo {author}
  {\bibfnamefont {P.}~\bibnamefont {Kuo}}, \bibinfo {author} {\bibfnamefont
  {D.~P.}\ \bibnamefont {Tsai}}, \ and\ \bibinfo {author} {\bibfnamefont
  {N.~I.}\ \bibnamefont {Zheludev}},\ }\href@noop {} {\bibfield  {journal}
  {\bibinfo  {journal} {Phys. Rev. Lett.}\ }\textbf {\bibinfo {volume} {104}},\
  \bibinfo {pages} {223901} (\bibinfo {year} {2010})}\BibitemShut {NoStop}%
\bibitem [{\citenamefont {Adamo}\ \emph {et~al.}(2012)\citenamefont {Adamo},
  \citenamefont {Ou}, \citenamefont {So}, \citenamefont {Jenkins},
  \citenamefont {{De Angelis}}, \citenamefont {{MacDonald}}, \citenamefont {{Di
  Fabrizio}}, \citenamefont {Ruostekoski},\ and\ \citenamefont
  {Zheludev}}]{Adamo2012}%
  \BibitemOpen
  \bibfield  {author} {\bibinfo {author} {\bibfnamefont {G.}~\bibnamefont
  {Adamo}}, \bibinfo {author} {\bibfnamefont {J.~Y.}\ \bibnamefont {Ou}},
  \bibinfo {author} {\bibfnamefont {J.~K.}\ \bibnamefont {So}}, \bibinfo
  {author} {\bibfnamefont {S.~D.}\ \bibnamefont {Jenkins}}, \bibinfo {author}
  {\bibfnamefont {F.}~\bibnamefont {{De Angelis}}}, \bibinfo {author}
  {\bibfnamefont {K.~F.}\ \bibnamefont {{MacDonald}}}, \bibinfo {author}
  {\bibfnamefont {E.}~\bibnamefont {{Di Fabrizio}}}, \bibinfo {author}
  {\bibfnamefont {J.}~\bibnamefont {Ruostekoski}}, \ and\ \bibinfo {author}
  {\bibfnamefont {N.~I.}\ \bibnamefont {Zheludev}},\ }\href@noop {} {\bibfield
  {journal} {\bibinfo  {journal} {Phys. Rev. Lett.}\ }\textbf {\bibinfo
  {volume} {109}},\ \bibinfo {pages} {217401} (\bibinfo {year}
  {2012})}\BibitemShut {NoStop}%
\bibitem [{\citenamefont {Meir}\ \emph {et~al.}(2014)\citenamefont {Meir},
  \citenamefont {Schwartz}, \citenamefont {Shahmoon}, \citenamefont {Oron},\
  and\ \citenamefont {Ozeri}}]{Meir2013}%
  \BibitemOpen
  \bibfield  {author} {\bibinfo {author} {\bibfnamefont {Z.}~\bibnamefont
  {Meir}}, \bibinfo {author} {\bibfnamefont {O.}~\bibnamefont {Schwartz}},
  \bibinfo {author} {\bibfnamefont {E.}~\bibnamefont {Shahmoon}}, \bibinfo
  {author} {\bibfnamefont {D.}~\bibnamefont {Oron}}, \ and\ \bibinfo {author}
  {\bibfnamefont {R.}~\bibnamefont {Ozeri}},\ }\href {\doibase
  10.1103/PhysRevLett.113.193002} {\bibfield  {journal} {\bibinfo  {journal}
  {Phys. Rev. Lett.}\ }\textbf {\bibinfo {volume} {113}},\ \bibinfo {pages}
  {193002} (\bibinfo {year} {2014})}\BibitemShut {NoStop}%
\bibitem [{\citenamefont {Brandes}(2005)}]{Brandes2005}%
  \BibitemOpen
  \bibfield  {author} {\bibinfo {author} {\bibfnamefont {T.}~\bibnamefont
  {Brandes}},\ }\href@noop {} {\bibfield  {journal} {\bibinfo  {journal}
  {Physics Reports}\ }\textbf {\bibinfo {volume} {408}},\ \bibinfo {pages}
  {315} (\bibinfo {year} {2005})}\BibitemShut {NoStop}%
\bibitem [{\citenamefont {Pierrat}\ and\ \citenamefont
  {Carminati}(2010)}]{Pierrat2010}%
  \BibitemOpen
  \bibfield  {author} {\bibinfo {author} {\bibfnamefont {R.}~\bibnamefont
  {Pierrat}}\ and\ \bibinfo {author} {\bibfnamefont {R.}~\bibnamefont
  {Carminati}},\ }\href@noop {} {\bibfield  {journal} {\bibinfo  {journal}
  {Phys. Rev. A}\ }\textbf {\bibinfo {volume} {81}},\ \bibinfo {pages} {063802}
  (\bibinfo {year} {2010})}\BibitemShut {NoStop}%
\bibitem [{\citenamefont {Diniz}\ \emph {et~al.}(2011)\citenamefont {Diniz},
  \citenamefont {Portolan}, \citenamefont {Ferreira}, \citenamefont
  {G{\'e}rard}, \citenamefont {Bertet},\ and\ \citenamefont
  {Auff{\`e}ves}}]{Diniz2011}%
  \BibitemOpen
  \bibfield  {author} {\bibinfo {author} {\bibfnamefont {I.}~\bibnamefont
  {Diniz}}, \bibinfo {author} {\bibfnamefont {S.}~\bibnamefont {Portolan}},
  \bibinfo {author} {\bibfnamefont {R.}~\bibnamefont {Ferreira}}, \bibinfo
  {author} {\bibfnamefont {J.~M.}\ \bibnamefont {G{\'e}rard}}, \bibinfo
  {author} {\bibfnamefont {P.}~\bibnamefont {Bertet}}, \ and\ \bibinfo {author}
  {\bibfnamefont {A.}~\bibnamefont {Auff{\`e}ves}},\ }\href@noop {} {\bibfield
  {journal} {\bibinfo  {journal} {Phys. Rev. A}\ }\textbf {\bibinfo {volume}
  {84}},\ \bibinfo {pages} {063810} (\bibinfo {year} {2011})}\BibitemShut
  {NoStop}%
\bibitem [{\citenamefont {{Jackson}}(1998)}]{Jackson}%
  \BibitemOpen
  \bibfield  {author} {\bibinfo {author} {\bibfnamefont {J.~D.}\ \bibnamefont
  {{Jackson}}},\ }\href@noop {} {\emph {\bibinfo {title} {Classical
  Electrodynamics}}}\ (\bibinfo  {publisher} {John Wiley \& Sons},\ \bibinfo
  {address} {New York},\ \bibinfo {year} {1998})\BibitemShut {NoStop}%
\bibitem [{\citenamefont {Born}\ and\ \citenamefont {Wolf}(1999)}]{Born}%
  \BibitemOpen
  \bibfield  {author} {\bibinfo {author} {\bibfnamefont {M.}~\bibnamefont
  {Born}}\ and\ \bibinfo {author} {\bibfnamefont {E.}~\bibnamefont {Wolf}},\
  }\href@noop {} {\emph {\bibinfo {title} {Principles of Optics}}},\ \bibinfo
  {edition} {7th}\ ed.\ (\bibinfo  {publisher} {Cambridge University Press,
  Cambridge, UK},\ \bibinfo {year} {1999})\BibitemShut {NoStop}%
\bibitem [{\citenamefont {Friedberg}\ \emph {et~al.}(1973)\citenamefont
  {Friedberg}, \citenamefont {Hartmann},\ and\ \citenamefont
  {Manassah}}]{Friedberg1973}%
  \BibitemOpen
  \bibfield  {author} {\bibinfo {author} {\bibfnamefont {R.}~\bibnamefont
  {Friedberg}}, \bibinfo {author} {\bibfnamefont {S.~R.}\ \bibnamefont
  {Hartmann}}, \ and\ \bibinfo {author} {\bibfnamefont {J.~T.}\ \bibnamefont
  {Manassah}},\ }\href@noop {} {\bibfield  {journal} {\bibinfo  {journal}
  {Physics Report}\ }\textbf {\bibinfo {volume} {7}},\ \bibinfo {pages} {101}
  (\bibinfo {year} {1973})}\BibitemShut {NoStop}%
\bibitem [{\citenamefont {Bourgain}\ \emph {et~al.}(2013)\citenamefont
  {Bourgain}, \citenamefont {Pellegrino}, \citenamefont {Fuhrmanek},
  \citenamefont {Sortais},\ and\ \citenamefont {Browaeys}}]{Bourgain2013b}%
  \BibitemOpen
  \bibfield  {author} {\bibinfo {author} {\bibfnamefont {R.}~\bibnamefont
  {Bourgain}}, \bibinfo {author} {\bibfnamefont {J.}~\bibnamefont
  {Pellegrino}}, \bibinfo {author} {\bibfnamefont {A.}~\bibnamefont
  {Fuhrmanek}}, \bibinfo {author} {\bibfnamefont {Y.~R.~P.}\ \bibnamefont
  {Sortais}}, \ and\ \bibinfo {author} {\bibfnamefont {A.}~\bibnamefont
  {Browaeys}},\ }\href@noop {} {\bibfield  {journal} {\bibinfo  {journal}
  {Phys. Rev. A}\ }\textbf {\bibinfo {volume} {88}},\ \bibinfo {pages} {023428}
  (\bibinfo {year} {2013})}\BibitemShut {NoStop}%
\bibitem [{Note1()}]{Note1}%
  \BibitemOpen
  \bibinfo {note} {The pulse length is then increased to $700$\protect \tmspace
  +\thinmuskip {.1667em}ns to keep the integration time
  reasonable.}\BibitemShut {Stop}%
\bibitem [{Note2()}]{Note2}%
  \BibitemOpen
  \bibinfo {note} {Note that the density drops during the $2\protect \tmspace
  +\thinmuskip {.1667em}\mu $s excitation pulse and that this drop is
  particularly significant for short free-flights.}\BibitemShut {Stop}%
\bibitem [{Note3()}]{Note3}%
  \BibitemOpen
  \bibinfo {note} {See Supplemental Material}\BibitemShut {NoStop}%
\bibitem [{Note4()}]{Note4}%
  \BibitemOpen
  \bibinfo {note} {
  The shift can be recovered in the low atom densities or when a specific collective mode that exhibits a shift is
  driven~\cite{Javanainen2016,Roof16,Araujo16}}\BibitemShut {NoStop}%
\bibitem [{\citenamefont {Roof}\ \emph {et~al.}(2016)\citenamefont {Roof},
  \citenamefont {Kemp}, \citenamefont {Havey},\ and\ \citenamefont
  {Sokolov}}]{Roof16}%
  \BibitemOpen
  \bibfield  {author} {\bibinfo {author} {\bibfnamefont {S.}~\bibnamefont
  {Roof}}, \bibinfo {author} {\bibfnamefont {K.}~\bibnamefont {Kemp}}, \bibinfo
  {author} {\bibfnamefont {M.}~\bibnamefont {Havey}}, \ and\ \bibinfo {author}
  {\bibfnamefont {I.}~\bibnamefont {Sokolov}},\ }\href@noop {} {\enquote
  {\bibinfo {title} {eprint arxiv:1603.07268},}\ } (\bibinfo {year}
  {2016})\BibitemShut {NoStop}%
\bibitem [{\citenamefont {Ara\'ujo}\ \emph {et~al.}(2016)\citenamefont
  {Ara\'ujo}, \citenamefont {Kre\v{s}i\'c}, \citenamefont {Kaiser},\ and\
  \citenamefont {Guerin}}]{Araujo16}%
  \BibitemOpen
  \bibfield  {author} {\bibinfo {author} {\bibfnamefont {M.}~\bibnamefont
  {Ara\'ujo}}, \bibinfo {author} {\bibfnamefont {I.}~\bibnamefont
  {Kre\v{s}i\'c}}, \bibinfo {author} {\bibfnamefont {R.}~\bibnamefont
  {Kaiser}}, \ and\ \bibinfo {author} {\bibfnamefont {W.}~\bibnamefont
  {Guerin}},\ }\href@noop {} {\enquote {\bibinfo {title} {eprint
  arxiv:1603.07204},}\ } (\bibinfo {year} {2016})\BibitemShut {NoStop}%
\end{thebibliography}
\end{document}